\documentstyle[aps,preprint]{revtex}

\begin{document}

\newcommand{\bqq}{\begin{equation}}
\newcommand{\eqq}{\end{equation}}

\title{Third order dielectric susceptibility in a model quantum paraelectric}

\author{R. Marto\v{n}\'{a}k\cite{byline}$^{a,b,c}$, E. Tosatti$^{b,c,d}$}

\address{$^a$Johannes-Gutenberg Universit\"{a}t, Institut f\"{u}r Physik, 
WA 331, \\
D-55099 Mainz, Germany \\
$^b$International Centre for Theoretical Physics (ICTP), \\
P.O. Box 586, 34014 Trieste, Italy\\
$^c$International School for Advanced Studies (S.I.S.S.A.),\\
Via Beirut 2--4, 34014 Trieste, Italy \\
$^d$Istituto Nazionale Fisica della Materia,\\
Unit\`{a} Trieste SISSA, Italy}

\maketitle

\begin{abstract}
\noindent
In the context of perovskite quantum paraelectrics, we study the effects of 
a quadrupolar interaction $J_q$, in addition to the standard dipolar one $J_d$.
We concentrate here on the nonlinear dielectric response $\chi_{P}^{(3)}$, as 
the main 
response function sensitive to quadrupolar (in our case antiquadrupolar) 
interactions. We employ a 3D quantum four-state lattice model and mean-field 
theory. The results show that inclusion of quadrupolar coupling of moderate 
strength ($J_q \sim {{1}\over{4}} J_d$) is clearly accompanied by a double 
change of sign of $\chi_{P}^{(3)}$ from negative to positive, near the quantum 
temperature $T_Q$
where the quantum paraelectric behaviour sets in. We fit our $\chi_{P}^{(3)}$
to recent experimental data for SrTiO$_3$, where the sign change is identified 
close to $T_Q \sim 37 K$.  
\end{abstract}

\section{Introduction}

Classical perovskite ferroelectrics such as BaTiO$_3$ are very widely studied.
As is well brought out by recent quantitative ab-initio studies
\cite{cohen,resta,vanderbilt}, formation of a local dipole moment corresponds
to formation of a shorter, partially covalent Ti-O bond while dipole-dipole 
interaction is partly of Coulomb origin (long-range) and partly resulting from 
the overlap of atomic orbitals (short-range).

The situation is 
considerably more open and interesting in those cases, mainly SrTiO$_3$ (also
KTaO$_3$) where quantum fluctuations are important to the extent that they  
actually suppress ferroelectricity altogether ("quantum paraelectrics" 
- M\"{u}ller and Burkhard \cite{mb}). 
Experimentally, the onset of quantum paraelectric regime in SrTiO$_3$ at 
a "quantum temperature" $T_Q \sim 30 \div 40$ K has revealed a variety of 
intriguing features. 
They are connected in particular with a surprisingly abrupt off-center local
displacement of the transition metal ion, without ferroelectric or 
structural long-range order, leading to phase transition-like features in 
EPR \cite{mbt}, EXAFS \cite{maglione}, X-ray \cite{neumann} and 
sound velocity \cite{fossheim,balashova,courten}. A similarly abrupt phenomenon
has been observed also in NMR studies of KTaO$_3$ \cite{rod}. 
In SrTiO$_3$, various spectroscopic anomalies have
also been found in Raman and Brillouin studies \cite{courtens}.

The reason why quantum fluctuations are so important here is not because the 
ions are particularly light. Rather, it is because the lattice packing is so 
tight, as compared, e.g. to BaTiO$_3$,  to leave very little room for the Ti 
ion to move off-center and form a preferential Ti-O dipole. As the lattice is 
compressed, the classical ferroelectric Ti off-center equilibrium displacement 
gets smaller and smaller, and the system approaches the classical displacive 
limit, characterized by a vanishing classical Curie temperature $T_c$.
In SrTiO$_3$, the extrapolated classical off-center Ti displacement and Curie
temperature are $\sim 0.03 \AA$ and $\sim$ 37 K respectively, against 
$\sim 0.1 \AA$ and $\sim$ 400 K in BaTiO$_3$.
In such a situation, quantum fluctuations, even if due to a heavy ion such as 
Ti, can well remove the ferroelectric long-range order and cause a persistence 
of the paraelectric state down to $T \rightarrow 0$. We have previously 
discussed this 
scenario in some detail \cite{tmssc} and also provided a quantum Monte Carlo 
study of a lattice model illustrating this phenomenon \cite{rm}.
A very recent ab-initio quantum Monte Carlo study of SrTiO$_3$ and BaTiO$_3$ has moreover
indicated that the exquisite sensitivity of the ferroelectric order parameter
to this kind of quantum fluctuations is likely to be related to the very 
anisotropic, quasi-one-dimensional electric dipole correlations, which are
absent for other "neutral" structural order parameters \cite{zhong}.

The theoretical study of structural phase transitions and ferroelectricity 
has so far mostly been conducted on models with an anharmonic on-site 
potential and a bilinear intercell interaction (standard model of structural 
phase transition, \cite{bruce,lines}), as well as on more specific shell models
\cite{bilz,migoni}.
Likewise, the studies of quantum paraelectricity have been performed within this
model \cite{oppermann,schneider}, or some approximate version of it, like the 
Ising model in transverse field \cite{kolb} or the quantum 4-state clock model 
\cite{rm}.
All these studies predict the existence of a critical value for the strength 
of quantum fluctuations, above which static ferroelectricity disappears in the 
ground state. The 
crossover from the high temperature classical paraelectric to low temperature
quantum paraelectric is predicted to be a smooth crossover, without any sharp
features. These models are therefore unable to account for the experimentally 
observed abrupt phenomena at $T_Q \sim 30 \div 40$ K in SrTiO$_3$.

While a thorough understanding of what is going on is still missing, it seems 
clear
that on cooling below $T_Q$ the individual Ti ions move off-center inside their
cell with surprising abruptness, whereas their mutual ordering between different
cells only progresses slowly as $T$ is lowered and never becomes total, down to
the lowest temperatures \cite{tmssc}. An abrupt onset of the local ferroelectric
lattice distortion
is {\it not} well explained by the standard model even if the local coupling of 
the polarization to the lattice strain is 
taken into account, as done for example by Zhong and Vanderbilt \cite{zhong}. 
In the latter quantum study, 
no particular sharp features have been observed on cooling the system down to 
T $\sim$ 5 K, contrary to the experimental observations. It therefore appears
highly probable that in order to achieve a satisfactory description of low
temperature regime of SrTiO$_3$, the extension of the standard model to include
some physically new ingredient is required.

A very natural extension is, apart from introducing a
coupling of the soft mode to other phonon modes, to drop the limitation of 
a simple, dipolar interaction $\cos(\phi_i - \phi_j)$ and take into account 
also higher angular terms ($\phi_i$ is the phase angle of a 2-component XY 
order parameter in cell $i$, see section 2).
The next order term with square symmetry has an angular dependence 
$\cos 2(\phi_i - \phi_j)$, typical of the 
{\it quadrupolar} interactions. \footnote{This is not the full form of the true
Coulomb quadrupole-quadrupole interaction, much as $\cos(\phi_i - \phi_j)$ is
not that of a true dipole-dipole interaction. These forms are adopted here for 
their simplicity.}

This paper is devoted to a study of the effects of such a quadrupolar-type of
interaction on the onset of quantum paraelectricity in SrTiO$_3$.
Deferring to section 2 a more detailed presentation of a simplified lattice 
hamiltonian which takes both dipolar and quadrupolar coupling into account, we 
note that the impact of the new coupling might be of qualitative importance.
In fact, as we pointed out in 
\cite{tmssc}, quantum effects in a quadrupolar system are very different from 
those in a dipolar system. A {\it purely} quadrupolar quantum
system exhibits {\it reentrance} in its phase diagram, while a dipolar one does
not. Reentrance in quantum quadrupolar systems is beautifully exemplified by the
data of Silvera et al. on ordering of HD under pressure \cite{silvera}. 
It also appears very straightforwardly in
simple models, at least in the mean-field theory \cite{tmssc,sumarokov}.

The ferroelectric perovskites, unlike HD, are unlikely to be dominated by the 
quadrupolar interaction. Our point, however, will be that they
exhibit features which cannot be explained by purely dipolar couplings.
Experimentally, direct detection 
of quadrupolar couplings in a ferroelectric system is not completely
straightforward, since the main quantity, the linear dielectric susceptibility
$\chi_1$, is unaffected by them, at least on the mean-field level \cite{morin}.
However, the third order nonlinear dielectric susceptibility $\chi_{P}^{(3)}$, 
defined 
by $P = \chi_{P}^{(1)} E + \chi_{P}^{(3)} E^3 + \ldots$, turns out to be very sensitive to 
quadrupolar couplings.
In particular Morin and Schmitt \cite{morin} (in a completely parallel magnetic
context) showed that $\chi_{P}^{(3)}$ will change sign from negative to 
positive, as a
system reaches a temperature region where quadrupolar effects become important,
while it will stay negative so long as they are unimportant. 

Very recently, new detailed data have been obtained by Hemberger, Lunkenheimer,
Viana, B\"{o}hmer and Loidl \cite{loidl}, who measured $\chi_{P}^{(3)}$ in 
SrTiO$_3$. 
They find strong structures in $\chi_{P}^{(3)}$ between 30 and 60 K, precisely 
in the region near $T_Q$, where the Ti ions suddenly move off-center on cooling,
giving rise to quantum fluctuating local dipoles, and quantum paraelectricity 
sets in.

In this paper we study this problem, and show that the behaviour of
$\chi_{P}^{(3)}$ can be understood if we assume the existence of quadrupolar 
interactions in SrTiO$_3$, besides the standard dipolar ones. In section 2, we 
first introduce a simple 
quantum lattice model including dipolar as well as and quadrupolar couplings.
In section 3 we calculate $\chi_{P}^{(3)}$ within mean-field theory, and find 
that depending on
the relative strength of the quadrupolar and dipolar interactions, 
$\chi_{P}^{(3)}$
exhibits a rich behaviour as a function of temperature. In section 4 we consider
the application of this calculation to the case of SrTiO$_3$. By comparing the 
calculated temperature dependence of the third order dielectric susceptibility 
to the observed data \cite{loidl} we estimate the strength of effective 
quadrupolar interactions in SrTiO$_3$. 
The results indicate that in SrTiO$_3$ a double change of sign of 
$\chi_{P}^{(3)}$ should take place, above and below $T \sim T_Q$, with 
decreasing temperature. The upper 
one from negative to positive $\chi_{P}^{(3)}$ is related to onset of 
quadrupolar 
effects; the lower one from positive back to negative $\chi_{P}^{(3)}$, to the 
onset of quantum paraelectricity. Before concluding, in section 5, we shall 
briefly discuss the possible origin of the quadrupolar
interactions in connection with the lattice dynamics of SrTiO$_3$. 

\section{Simplified lattice model of {\mbox SrTiO$_3$}}

The hamiltonian is quite generally a function of $3 \times 5 = 15$ continuous 
coordinates per cell in the cubic phase (5 atoms per unit cell) or of even
more coordinates in the tetragonal antiferrodistorted state.
Even simplifying to include only polarization, strain and antiferrodistortive 
degrees of freedom, we are still forced to include at least $3 + 3 + 3 = 9$ 
degrees of freedom per cell. A proper quantum treatment, for example via
PIMC \cite{zhong} of so many continuous degrees of freedom is at present
very demanding. A quantum lattice model, although much less realistic, and 
incapable of describing the displacive behaviour of a perovskite in its
classical regime ($T \gg T_Q$) is more directly amenable to study either via 
PIMC, or, much more simply, via mean-field theory.  

In this section we introduce a simplified discrete lattice model of a QPE system
including dipolar as well as quadrupolar interactions.
A discussion of these simple models in the context of quantum paraelectricity
of SrTiO$_3$ has already been presented in Ref.\cite{rm,tmssc}, where
a variety of lattice models has been introduced and considered. For convenience 
we briefly repeat here the main ideas.
	
We start from the experimentally observed XY-character of the incipient
ferroelectricity in tetragonal SrTiO$_3$. The physical reason why at zero stress
and field the polarization is confined to the $(x,y)$ plane is given by the 
presence (below 105 K) of the antiferrodistortive order parameter, which 
expands the lattice in $(x,y)$ plane, while contracting it along the $z$-axis,
taken along (001). 
The onset of ferroelectric polarization is favored in the expanded $(x,y)$
plane, and disfavoured in the contracted $z$-direction. In the $(x,y)$ plane,
the polarization has four easy directions, namely $(\pm100),(0\pm10)$. The 
corresponding quantum four-state clock model introduced first in \cite{rm} and 
considered in various versions in \cite{tmssc} appears thus to be a 
minimal model for the description of SrTiO$_3$ at sufficiently low temperatures.
The model neglects the radial degrees of freedom associated with the continuous
displacement of the Ti ion from the center of the oxygen cage in the cell $i$ of
the crystal. It assumes that the displacement is of fixed magnitude and can be 
completely characterized by a discrete plane rotor angular variable $\phi_i$. 
This variable is allowed four possible values
$\phi_i = 0, \pi/2, \pi, -\pi/2$, and the corresponding four quantum states 
labeled as $|0\rangle,|1\rangle,|2\rangle,|3\rangle$ constitute a basis of the 
on-site Hilbert space. The quantum effects are mimicked by allowing the clock 
variable $\phi_{i}$ to hop onto its two nearest orientations, i.e. from 
$\phi_{i}$ into $\phi_{i} \pm {{\pi}\over{2}}$ with amplitude $-t$. This is 
expressed by a 
hamiltonian term $H^{hop}_i$, represented in the site basis by the matrix
\bqq
H^{hop}_{i} = t \left|
\begin{array}{cccc}
0 & -1 & 0 & -1 \\
-1 & 0 & -1 & 0 \\
0 & -1 & 0 & -1  \\
-1 & 0 & -1 & 0  
\end{array} \right| \; . \label{mat}
\eqq 
For simplicity, interaction between different cells $i \neq j$ is limited to 
nearest 
neighbours, and the adjacent cells are supposed to interact via the usual 
"dipole" interaction
\bqq
H^{d} = - J_{d} \sum_{\langle ij \rangle} \cos (\phi_{i} - \phi_{j}) \,, 
\label{dip}
\eqq
as well as via a "quadrupole" interaction  
\bqq
H^{q} = - J_{q} \sum_{\langle ij \rangle} \cos 2 (\phi_{i} - \phi_{j}) \; .
\label{quad}
\eqq
While the origin of $H^{d}$ is widely discussed, 
that of $H^{q}$ is not, and this is to our knowledge the first place where such
a term is invoked for a displacive ferroelectric. The strength
of the $H^{q}$ term in SrTiO$_3$ will be estimated in section 4 and its possible
origin will be discussed in section 5. We note that the {\it classical}
four-state model $H = H^{d} + H^{q}$ is the well-known Ashkin-Teller model 
\cite{baxter}. The 2D classical A-T model is already considerably richer than the
XY model (\ref{dip}). For the sake of simplicity, we shall treat our model on 
a 3D cubic lattice, and use a mean-field theory. This approximation, while far 
from accurate, is at least not totally unacceptable in (3+1)D and is attractive
due to its great simplicity.
 
The complete hamiltonian that will be considered in the following is then given
by 
\bqq
H = \sum_i H^{hop}_{i} + H^d + H^q \; .
\label{ham}
\eqq

\section{Mean field theory of third order dielectric susceptibility}

In this section we start from our microscopic hamiltonian (\ref{ham}) and derive
a free energy expansion, using a mean-field approximation. 
From the free energy expansion we then calculate the third order dielectric 
susceptibility $\chi_P^{(3)}$ as a function of temperature. Our treatment will 
parallel very closely that of Morin \& Schmitt \cite{morin} for the magnetic 
case. We will show that the behaviour of $\chi_P^{(3)}$ at low and intermediate
temperatures relative to the quantum temperature $T_Q$ is 
profoundly modified by the presence of the quadrupolar interactions.

We take as our mean-field hamiltonian a sum of on-site mean-field hamiltonians, 
\bqq
H^{0} = \sum_i H_{i}^{0} = \sum_i (H_{i}^{hop} - e P_{xi} - q R_{i}) \; , 
\label{mfham}
\eqq
corresponding to a single quantum rotor in external fields $e$ and $q$,
coupling to the on-site polarization and quadrupole moment $P_{xi}$ and $R_i$, 
respectively. For simplicity, we shall assume from now on that 
the polarization component $P_y$ is always zero and omit the subscript $x$ on
$P_x$; we also set $|t|=1$ as the energy and temperature scale. The matrix of
the on-site hamiltonian (\ref{mfham}) in the basis of the 4 states 
$|1\rangle,|2\rangle,|3\rangle,|4\rangle$ reads
\bqq
H^{0}_{i} = \left|
\begin{array}{cccc}
q & -1 & 0 & -1 \\
-1 & e - q & -1 & 0 \\
0 & -1 & q & -1  \\
-1 & 0 & -1 & -e - q  
\end{array} \right| \; . \label{mfmat}
\eqq

	The trial free energy $F_t(e,q)$ per site as a function of the 
mean-field variational parameters $e,q$ is given by 
\bqq
F_{t}(e,q) = F_{0} + \langle H - H^{0} \rangle_0 = F_0 - 6 J_d {{1}\over{2}} 
P^2 - 6 J_q {{1}\over{2}} R^2 + e P + q R \; , \label{ft0}
\eqq
where
\bqq
F_{0} = -{{1}\over{\beta}} {\mbox Tr} \; e^{-\beta H^{0}} \; , \label{f0}
\eqq
and $P,R$ are now a ferroelectric and a quadrupolar order parameter.
The hamiltonian (\ref{mfmat}) cannot be diagonalized analytically, but the 
expansions of the energy levels in powers of the fields $e,q$ around their zero 
values (we assume that our system, like SrTiO$_3$ in the absence of fields, is 
paraelectric) can be found by perturbation theory. For our
purpose it is sufficient to find the expansion to fourth order in $e$ and to 
second order in $q$. The following expressions are found for the eigenvalues of
$H_i^{0}$
\begin{eqnarray}
E_0 &=& -2 - {{1}\over{4}} e^2 + {{1}\over{64}} e^4 - ({{1}\over{4}} +
{{3}\over{32}} e^2) q^2 + {{1}\over{4}} e^2 q + \ldots \\
E_1 &=& -q \\
E_2 &=& (1 - {{1}\over{2}} e^2) q + \ldots \\
E_3 &=& 2 + {{1}\over{4}} e^2 - {{1}\over{64}} e^4 + ({{1}\over{4}} +
{{3}\over{32}} e^2) q^2 + {{1}\over{4}} e^2 q + \ldots 
\end{eqnarray}
whence the power expansion of free energy (\ref{f0}) can be calculated,
\bqq
F_0 = -{{1}\over{2}} \chi_{0}^{(1)} e^2 - {{1}\over{2}} \chi_2 q^2 - 
\chi_{2}^{(2)} e^2 q - {{1}\over{4}} \chi_{0}^{(3)} e^4 + \ldots \; , 
\label{f0exp}
\eqq 
where 
\begin{eqnarray}
\chi_{0}^{(1)} &=& {{1}\over{2}} \tanh \beta  \nonumber \\
\chi_{0}^{(3)} &=& {{1 + 4 \beta e^{2 \beta} - e^{4 \beta}}
\over{16 (1 + e^{2 \beta})^2}} \nonumber \\
\chi_{2} &=& - {{1 - 4 \beta e^{2 \beta} - e^{4 \beta}}
\over{2 (1 + e^{2 \beta})^2}} \nonumber \\
\chi_{2}^{(2)} &=& -{{1}\over{4}} \tanh^2 \beta \; . \label{chi}
\end{eqnarray}

The polarization and quadrupolar order parameters $P$ and $R$ can now be 
calculated as derivatives of 
(\ref{f0exp}) with respect to the fields $e$ and $q$
\begin{eqnarray}
P &=& - {{\partial F_0} \over {\partial e}} = \chi_{0}^{(1)} e + 
2 \chi_{2}^{(2)} e q + \chi_{0}^{(3)} e^3 + \ldots \label{op0} \\
R &=& - {{\partial F_0} \over {\partial q}} = 
\chi_{2} q + \chi_{2}^{(2)} e^2 + \ldots \; . \label{op}
\end{eqnarray}
	In order to obtain an expansion of the free energy term $F_0$ in terms 
of the order parameters $P,R$, we must invert the expansions (\ref{op0}) and
(\ref{op}) and express the fields $e,q$ in terms of powers of $P$ and $R$. We 
find the expressions 
\begin{eqnarray}
e &=& {{1}\over{\chi_0^{(1)}}} P - {{2 \chi_2^{(2)}}\over{(\chi_0^{(1)})^2 
\chi_2}} P R  - {{1}\over{(\chi_0^{(1)})^4}} \left(\chi_0^{(3)} - 
{{2 (\chi_2^{(2)})^2}\over {\chi_2}} \right) P^3 + \ldots \\
q &=& {{1}\over{\chi_2}} R - {{\chi_2^{(2)}}\over{(\chi_0^{(1)})^2 
\chi_2}} P^2 + \ldots \; , 
\end{eqnarray}
which after substituting to (\ref{f0exp}) yields the desired form 
\bqq
F_0 = - {{1}\over{2}} (\chi_{0}^{(1)})^{-1} P^2 - {{1}\over{2}} 
(\chi_{2})^{-1} R^2 + 
{{2 \chi_2^{(2)}}\over{(\chi_0^{(1)})^2 \chi_2}} P^2 R +
\left( {{3}\over{4}} {{\chi_0^{(3)}}\over{(\chi_0^{(1)})^4}} - 
{{3}\over{2}} {{(\chi_2^{(2)})^2}\over{\chi_2 (\chi_0^{(1)})^4}} 
\right) P^4 + \ldots\; . \label{ff0}
\eqq

At this point, we can express the full free energy $F_t$ (\ref{ft0}) as an 
expansion in powers of the order parameters, namely polarization $P$ and 
quadrupole moment $R$. It reads
\begin{eqnarray}
F_t(P,R) &=& {{1}\over{2}} [(\chi_{0}^{(1)})^{-1} - 6 J_d] P^2 + 
{{1}\over{2}} [(\chi_{2})^{-1} - 6 J_q] R^2 - 
{{\chi_2^{(2)}}\over{(\chi_0^{(1)})^2 \chi_2}} P^2 R \nonumber \\
&-& {{1}\over{4}} {{1}\over{(\chi_0^{(1)})^4}} \left( \chi_0^{(3)} -
2 {{(\chi_2^{(2)})^2}\over{\chi_2}} \right) P^4 \; . \label{ft}
\end{eqnarray}

As in real SrTiO$_3$ at zero stress and field, we assume $P = R = 0$. 
In the neighbourhood of this minimum of $F_t$, we now wish to
eliminate $R$, yielding an effective free energy expansion in powers 
of the polarization $P$ alone. Minimizing (\ref{ft}) with respect to $R$, we
obtain
\bqq
R = {{\chi_2^{(2)}}\over{(\chi_0^{(1)})^2 (1 - 6 J_q \chi_2)}} P^2 \; ,
\eqq
and after substituting this expression back to (\ref{ft}) we obtain the 
expansion of $F_t$ in terms of $P$ only 
\bqq
F_t(P) = {{1}\over{2}} [(\chi_{0}^{(1)})^{-1} - 6 J_d] P^2 - 
{{1}\over{4}} {{1}\over{(\chi_0^{(1)})^4}} \left( \chi_0^{(3)} +
{{12 J_q (\chi_2^{(2)})^2}\over{1 - 6 J_q \chi_2}} \right) P^4 \; . \label{ftp}
\eqq
Now it is worth noticing that the coefficient of $P^4$ has been renormalized 
downwards (i.e. towards negative values) by the quadrupolar coupling $J_q$. 
This follows because $\chi_{0}^{(3)}$ is negative definite, while
the other term dependent on $J_q$ is positive as long as $J_q < (6\chi_2)^{-1}$.

By adding an interaction term $- E P$, corresponding to an external electric 
field $E$, to the free energy (\ref{ftp}) and minimizing
with respect to the polarization $P$, we can  
calculate the first order (linear) and third order (nonlinear) dielectric 
susceptibilities $\chi_{P}^{(1)}$ and $\chi_{P}^{(3)}$, defined as usual by the
expansion
\bqq
P = \chi_{P}^{(1)} E + \chi_{P}^{(3)} E^3 + \ldots \; .
\eqq 
In our model we find
\begin{eqnarray}
\chi_{P}^{(1)} &=& {{\chi_0^{(1)}}\over{1 - 6 J_d \chi_0^{(1)}}} \label{chi1} \\
\chi_{P}^{(3)} &=& {{1}\over{(1 - 6 J_d \chi_0^{(1)})^4}} \left(\chi_0^{(3)} + 
{{12 J_q (\chi_2^{(2)})^2}\over{1 - 6 J_q \chi_2}} \right) \; . \label{chi3}
\end{eqnarray}
The nonlinear susceptibility (\ref{chi3}) can also be written in a slightly
more transparent form 
\bqq
\chi_{P}^{(3)} = \chi_P^{(3)}(J_q = 0) + 
{{12 J_q (\chi_P^{(1)})^4}\over{1 - 6 J_q \chi_2}} \; , \label{chi3p}
\eqq
where we made use of the relation $(\chi_{2}^{(2)})^2 = (\chi_{0}^{(1)})^4$, 
verified by Eq.(\ref{chi}). While the first order dielectric susceptibility 
$\chi_{P}^{(1)}$, in the mean field approximation, is independent
of $J_q$, we see that the third order susceptibility $\chi_{P}^{(3)}$ 
is a sum of its original value in the absence of quadrupolar coupling 
which is always {\it negative}, and of an additional 
{\it positive} term. The latter can eventually overbalance the negative term
and reverse the sign of $\chi_{P}^{(3)}$, if $J_q$ is large enough.
The temperature dependences of $\chi_{P}^{(3)}$ for $J_d = 0.273$ and a series 
of values of quadrupolar coupling $J_q$ are plotted on Fig.1. We see that 
a) the effect of renormalization of $\chi_{P}^{(3)}$ is most strongly 
pronounced at intermediate temperatures $T \sim T_Q$, and 
b) there is a region of
values of $J_q$ where $\chi_{P}^{(3)}$ now turns positive at intermediate 
temperatures, while staying negative at sufficiently high and low temperatures.
Finally, if $J_q$ is stronger, there is a 
temperature where $\chi_{P}^{(3)} \rightarrow \infty$ and an independent 
(anti)quadrupolar ordering takes place.

We point out here that positive $\chi_{P}^{(3)}$, which means a negative overall
coefficient in front of $P^4$ term in (\ref{ftp}) would imply the 
{\it ferroelectric} transition at the classical level to be of 
{\it first order}. The QPE state for positive $\chi_{P}^{(3)}$ can therefore 
be seen as a frustrated first order transition.

\section{Application to the case of SrTiO$_3$}

We now wish to apply the above results to the case of SrTiO$_3$. As mentioned 
in the introduction, SrTiO$_3$ belongs rather to the class of displacive than to
the order-disorder ferroelectrics. The use of a continuous model should thus be 
preferred to a discrete one, if a realistic comparison to experiment is to be 
attempted. However, the main purpose of this paper is to pursue the qualitative
differences induced in the behaviour of $\chi_{P}^{(3)}$ by a presence of 
a quadrupolar 
interaction with its different angular dependence, and these features should be 
correctly reproduced also by a discrete model, which in turn has the big 
advantage of being easily analytically tractable on a simple mean-field level. 
It is clearly beyond the possibilities of such a discrete model to describe the 
sharp dipole onset observed at
$T_Q \sim $37 K, since the local off-center displacement is described just by
the radial part of the local mode which is left out completely in the present 
discrete model.

Before attempting any comparison between experimental data \cite{loidl} and our 
results (\ref{chi1}) and (\ref{chi3}), we have to convert the latter formulas 
from dimensionless units back to normal ones. We consider first
the linear dielectric susceptibility. If the displacement of the
Ti ion from the center of the cage is associated with a dipole moment $\mu$,
the dipole density per unit volume being $n$, then $\chi_{P}^{(1)}$ can be
conventionally written in form of the familiar Barrett formula 
\bqq
\chi_{P}^{(1)} = { {{n \mu^2}\over{2 k_B \epsilon_0}} \over { {{T_1}\over{2}} 
\coth{{T_1}\over{2 T}} - T_0}} \; ,
\eqq
derived in many different contexts \cite{barrett,mb}. The temperatures 
$T_0$ and $T_1$ are given by $T_0 = 3 J_d/k_B, T_1 = 2 t/k_B$, and have the 
meaning of classical Curie temperature and of quantum temperature, respectively.
This formula has been used to fit the $\chi_{P}^{(1)}$ data on SrTiO$_3$ 
\cite{mb,loidl}, and produces a reasonable fit over a temperature range from
0 to 100 K with $T_1 \sim $ 88 K, $T_0 \sim $ 36 K, and 
${{n \mu^2}\over{2 k_B \epsilon_0}} \sim 10^5$ K. The microscopic hamiltonian 
parameters corresponding to these values of $T_0$ and $T_1$ are $J_d = $ 12 K, 
$t = $ 44 K. We note in particular that the ratio $J_d/t = 0.273$ is not 
far from the critical mean-field value of $(J_d/t)_c = 1/3$, where our model 
undergoes a transition from para ($J < J_c$) to ferro ($J > J_c$).

A problem arises, however, with the prefactor  
${{n \mu^2}\over{2 k_B \epsilon_0}}$, since with one dipole per cell 
$n = 1/a^3 = 1.68 \times 10^{28} m^{-3}$, where 
$a = 3.9 \AA$ is the lattice constant of SrTiO$_3$, the required value of the 
dipole moment per cell is $\mu = 3.4 \times 10^{-29}$ Cm. If we take the 
effective charge to be $Z^* \sim 8$ \cite{vanderbilt1}, the corresponding 
displacement should be $d \sim 0.26 \AA$, which is an order of magnitude larger
than the value of $\sim 0.03 \AA$ realistically estimated in Ref.\cite{mb},
as well as in ab-initio calculations 
\cite{vanderbilt}. The discrepancy is, in our opinion, related to the 
quantitative inadequacy of the discrete model to SrTiO$_3$,
as well as to the mean-field treatment. The Barrett formula 
reproduces the essential features of the temperature behaviour of the linear 
susceptibility well enough, but fails at the quantitative level.

Let us now come to the nonlinear susceptibility, and the related quadrupolar 
effects. The complete expression we have just obtained for $\chi_{P}^{(3)}$ is
\bqq
\chi_{P}^{(3)} = { {{n \mu^4 t}\over{\epsilon_0}} \over
{(t - 6 J_d \chi_0^{(1)})^4}} \left(\chi_0^{(3)} + 
{{12 J_q (\chi_2^{(2)})^2}\over{t - 6 J_q \chi_2}} \right) \; , \label{fchi3}
\eqq
where $\beta = t/(k_B T)$. We can now try 
to fit this formula to the $\chi_{P}^{(3)}$ data measured on SrTiO$_3$ 
\cite{loidl}. For this purpose, however, the form in which the data are 
presented in Ref. \cite{loidl}, namely $\log|\chi_{P}^{(3)}|$ vs. temperature
plot is not so well suited, since the mere knowledge of the magnitude of the 
complex quantity $|\chi_{P}^{(3)}|$ without its phase angle, mixes together the 
real and imaginary part and conceals the desired information about the sign of 
the real part. Our
static mean-field theory produces, of course, a {\it real} static
$\chi_{P}^{(3)}$. We have therefore attempted to extract Re $\chi_{P}^{(3)}$
from the data \cite{loidl}, making use of the corresponding phase angles 
kindly provided by J. Hemberger \cite{hemberger}. The resulting experimentally
derived Re $\chi_{P}^{(3)}$ (corresponding to the lowest electric field 
intensity $E_0 =$ 50 V/mm) is shown on Fig.2 as a function of temperature. 
While at low temperatures this quantity is negative, on heating it crosses zero
at $T = 33$ K where it becomes positive, passes through a maximum at $T = 45 $ K
and then approaches zero again, from the positive values. Above $T \sim 60$ K, 
it probably turns small and negative again, although experimental uncertainty 
appears larger in this regime. With hindsight, we re-examined also some old data
of Fleury and Worlock \cite{fleury} for SrTiO$_3$ and found that evidence for 
a positive $\chi_{P}^{(3)}$ at $T = 40$ K can be evinced there, too 
(Fig. 5 b, Ref. \cite{fleury}). 

	Having in mind the difficulties related to the prefactor, encountered
in case of $\chi_{P}^{(1)}$, we decided to allow this to be a free parameter.
Anticipating that the best fit of $\chi_{P}^{(3)}$ does not necessarily have to
lead to the same values of $J_d$ and $t$ as the fit of $\chi_{P}^{(1)}$, we 
decided to preserve the value of $J_d/t = 0.273$ and 
allow for $J_q$ and $t$ to be free parameters. The result of this fitting is 
the curve on Fig.2, which corresponds to $t = 75$ K, $J_d = 0.273 t = 20.5$ K 
and $J_q = 4.8$ K.
With these values, the positive maximum of $\chi_{P}^{(3)}$ is well fitted,
although at lower temperatures the fit is less perfect within the present 
theory.
At higher temperatures on the other hand it is hard to assess the agreement 
between the theory and data; in any case the theory predicts $\chi_{P}^{(3)}$ 
to become negative again at high temperatures and approach zero from below. 
From Fig.2 it is clear that by assuming a quadrupolar interaction 
of strength about 5 K, the theory can
qualitatively account for the temperature dependent double change of sign
of $\chi_{P}^{(3)}$. 
Quantitatively, the agreement is worse than in case of $\chi_{P}^{(1)}$, which
is perhaps not surprising for a thermodynamic quantity which is a higher 
order derivative of the free energy. The required value of $t$ is now larger by 
$\sim$ 70 \% and the value of the displacement required in order to obtain the 
necessary value of the prefactor is even an order of magnitude larger than 
in case of $\chi_{P}^{(1)}$, namely 2.3 $\AA$, which is clearly unphysical. 
As in the case of $\chi_{P}^{(1)}$, we attribute these discrepancies to the 
crudeness of the model. On the other hand, a firm result of this 
calculation and fit is that {\it the presence of $J_q$ appears to be crucial in
order to account for a positive $\chi_{P}^{(3)}$ at intermediate temperatures}.
 
\section{Possible origin of the quadrupolar interaction in perovskites}

Before closing, we discuss the possible microscopic origin of the effective
(anti)-quadrupolar interaction $J_q$ in perovskites. Recently, we did in fact 
invoke for SrTiO$_3$ \cite{tmssc} the possible importance of an effective
quadrupolar coupling $J_q$ between the dipoles in different cells, in
addition to the dipolar one $J_d$.  The sign of this 
interaction $J_q$ is negative ("antiquadrupolar"), such that dipoles prefer to 
be either parallel or antiparallel, to being orthogonal. One physical source
anticipated in Ref.\cite{tmssc}
could be related to a lattice strain effect, as follows. When a dipole bond 
appears in one cell, it will cause an elongation of that cell along the dipole
direction, and a shrinking in the orthogonal direction via electrostriction.
This deformation, to avoid strain gradients, will extend to the neighbouring 
cells where it will favor 
equally well a parallel or an antiparallel dipole orientation, but will 
disfavor the orthogonal one. The coupling then arises so as to reduce strain 
{\it gradients} associated with orthogonal dipoles, but absent for parallel or
antiparellel dipoles. In order to estimate the
strength of the resulting interaction, these intuitive considerations have to be
formalized.

Formally, the term representing the coupling between the local polarization
and strain is, for a perovskite structure, linear in strains and quadratic in 
polarizations, and according to Ref.\cite{vanderbilt} it reads
\bqq
H^{int} = {{1}\over{2}} \sum_i \sum_{l \alpha \beta} B_{l \alpha \beta}
\eta_l(\vec{R}_i) u_{\alpha}(\vec{R}_i) u_{\beta}(\vec{R}_i) \; \label{hint},
\eqq
while the elastic energy is given by
\bqq
H^{el} = {{1}\over{2}} \sum_i \sum_{l k} C_{l k} 
\eta_l(\vec{R}_i)  \eta_k(\vec{R}_i) \; , \label{hel}.
\eqq
where $\eta_l(\vec{R}_i)$ and $u_{\alpha}(\vec{R}_i)$ are the strain and 
local soft mode amplitude components at site $i$, respectively, 
$B_{l \alpha \beta}$ are the strain-ferroelectric soft mode coupling constants 
and 
$C_{l k}$ are the elastic constants. The total strain at site $i$ consists of 
the homogeneous component arising from the uniform deformation of the whole
system as well as of the inhomogeneous local strain. Both components have 
to be taken into account in a different way, because the 6 local strain 
components per 
cell are not in fact independent quantities, but rather suitable linear
combinations of the 3 independent acoustic displacement components per cell.
To proceed, one has to express the local strains in both (\ref{hint})
and (\ref{hel}) in terms of the acoustic displacements. Integrating out the 
acoustic displacements from the total hamiltonian, one finds the effective 
interaction between the polarizations in different cells, which
turns out to be fourth order in polarizations and contains also the 
quadrupolar interaction which we are interested in. 
In our case, however, such procedure would not be as straightforward 
as in the corresponding classical case, because the complete hamiltonian 
contains the quantum kinetic energy term which does not commute with the 
coupling term containing the polarization. The simpler classical calculation
should, however, provide at least a rough estimate of the strength of the
effective $J_q$ induced in this way. 

Before doing this calculation in detail, however, it is actually possible to 
estimate an upper limit to the strength of the effective $J_q$ by means of 
a simple argument. The induced {\it intersite} interaction term of the form
$(\vec{u}(\vec{R}_i) \cdot \vec{u}(\vec{R}_j))^2 = {{1}\over{2}} u(\vec{R}_i)^2 
u(\vec{R}_j)^2 (1 + \cos 2(\phi_i - \phi_j)), i \neq j$ certainly must be
weaker than an analogous induced term for $i = j$, 
$(\vec{u}(\vec{R}_i) \cdot \vec{u}(\vec{R}_i))^2 = u^4(\vec{R}_i)$, which 
represents a renormalization of the on-site quartic term $u^4(\vec{R}_i)$. 
This renormalization is clearly of the order of 
${{B_{l \alpha \beta}^2}\over{C_{l k}}}$. In order to get an upper estimate for
the interaction strength $J_q$ (\ref{quad}), we must also multiply by 
${{1}\over{2}} d^4$, where $d$ is the typical value of the local displacement. 
Taking the typical values of 
$B_{l \alpha \beta} \sim$ 1.4 hartree/bohr$^2$ and $C_{l k} \sim$ 5 hartree
(from Ref.\cite{vanderbilt}), 
together with the displacement $d \sim 0.03 \AA$, we arrive at a value about 
0.5 K, which is an order of magnitude lower than the value of 5 K 
estimated in the previous section to be necessary to explain $\chi_{P}^{(3)}$ 
in SrTiO$_3$.

A more accurate classical calculation of $J_q^{strain}$ is most conveniently 
performed in the Fourier representation, due to the translational invariance of 
the hamiltonian. The long-wavelength limit of the problem has been studied
in the 70's when it was shown that
the resulting interaction is nonanalytical in $k$-space and has a long-range 
tail in the real space \cite{bender,bruce}. A similar 2D calculation of indirect
strain-induced coupling has been performed for the case of SrTiO$_3$ 
\cite{rmun}, taking into account all the Fourier components from the Brillouin 
zone, which allows extraction of the 
induced interaction for each pair of lattice sites $i \neq j$. We will not
go here into the full detail of that calculation and merely quote the main
result, which is that the strength of the induced interaction is of the order
of $0.1$ K, a factor of five lower than the previous upper limit
estimate. We conclude that the origin of the effective quadrupolar interactions 
of the required strength is not likely to be elastic, and should be sought
elsewhere. 

Another possible source could be long-range interactions. Coupling 
one cell with further neighbours will generally yield terms with higher angular
Fourier components. The soft phonon mode dispersion away from $k = 0$ is in fact
very non-sinusoidal, precisely due to Coulomb interactions \cite{cochran}. 
A third possibility might finally arise from anharmonic terms like
$(\vec{u}(\vec{R}_i) \cdot \vec{u}(\vec{R}_j))^2$ of non-strain origin. At this 
stage we are not in a position to decide among these possibilities.
It would therefore be highly desirable to extend ab-initio studies similar to
those in Ref.\cite{vanderbilt} beyond the lowest order considered there
("local anharmonicity approximation"), 
and try to determine also the required higher order terms of the energy 
expansion.

Because of the close relationship between these expansions and the harmonic
and anharmonic phonon properties, which in case of KTaO$_3$ and KNbO$_3$ are
also quite well reproduced by extended shell models \cite{migoni1}, it might 
also be possible to use them for this purpose. 

\section{Conclusions} 

In conclusion, we have considered the nonlinear dielectric susceptibility
$\chi_{P}^{(3)}$, and its behaviour with temperature for a model system
exhibiting quantum paraelectricity below a "quantum temperature" $T_Q$. The 
model is chosen so as to include a quadrupolar coupling in addition to the 
normal
dipolar one. It is found that the effect of the quadrupolar couplings is to
cause a temporary switch of sign of $\chi_{P}^{(3)}$ from negative, well above
and well below $T_Q$, to positive at $T \sim T_Q$. Recent data for 
$\chi_{P}^{(3)}$ of SrTiO$_3$ appear to agree well with this predicted behaviour
implying an effective quadrupolar coupling $J_q \sim {{1}\over{4}} J_d$. Work 
which remains to be done includes\\
a) better understanding of the true microscopic origin of $J_q$, presently 
still unclear\\
b) understanding the possible relationship of this new coupling to the exciting
and largely unexplained phenomenology observed near $T_Q$, in M\"{u}ller's
original experiments \cite{mbt}, and those which followed it. For the time 
being, we have noted that a positive $\chi_{P}^{(3)}$ at $T_Q$ implies that the
QPE state results due to quantum frustration of an otherwise first-order
ferroelectric transition. 

\acknowledgements

We are grateful to A. Loidl, P. Lunkenheimer, R. B\"{o}hmer and J. Hemberger
for sharing with us their results prior to publication and especially to
J. Hemberger, for providing a copy of his Diploma Thesis \cite{dhem} 
and numerical values for the phase angle of $\chi_{P}^{(3)}$ \cite{hemberger}. 
One of us (R. M.) 
would also like to thank K. Binder, R. B\"{o}hmer, J. Hemberger, R. Hlubina, 
A. Loidl, P. Lunkenheimer and A. Milchev for interesting and stimulating 
discussions. We are especially grateful to R. Migoni for offering his insight 
in the ferroelectric perovskites. R. M. would also like to acknowledge the 
fellowship of the Max Planck Intitut f\"{u}r Polymerforschung, Mainz, Germany,
as well as the hospitality provided by SISSA and ICTP, Trieste, Italy.
Work at SISSA/ICTP was partly sponsored under EEC contracts ERBCHRXCT920062 and
ERBCHRXCT940438.

\begin{figure}
Fig.1. Third order nonlinear dielectric susceptibility $\chi_{P}^{(3)}$ as 
a function of temperature for a fixed value of dipolar coupling $J_d$ and 
various values of quadrupolar coupling $J_q$. 
\end{figure}

\begin{figure}
Fig.2. Fitting of the experimentally measured ${\mbox Re}(\chi_{P}^{(3)})$  
\cite{loidl,hemberger,dhem} of 
SrTiO$_3$ with formula (\ref{fchi3}). Full points correspond to experimental 
data, solid line to the theoretical curve.
\end{figure}

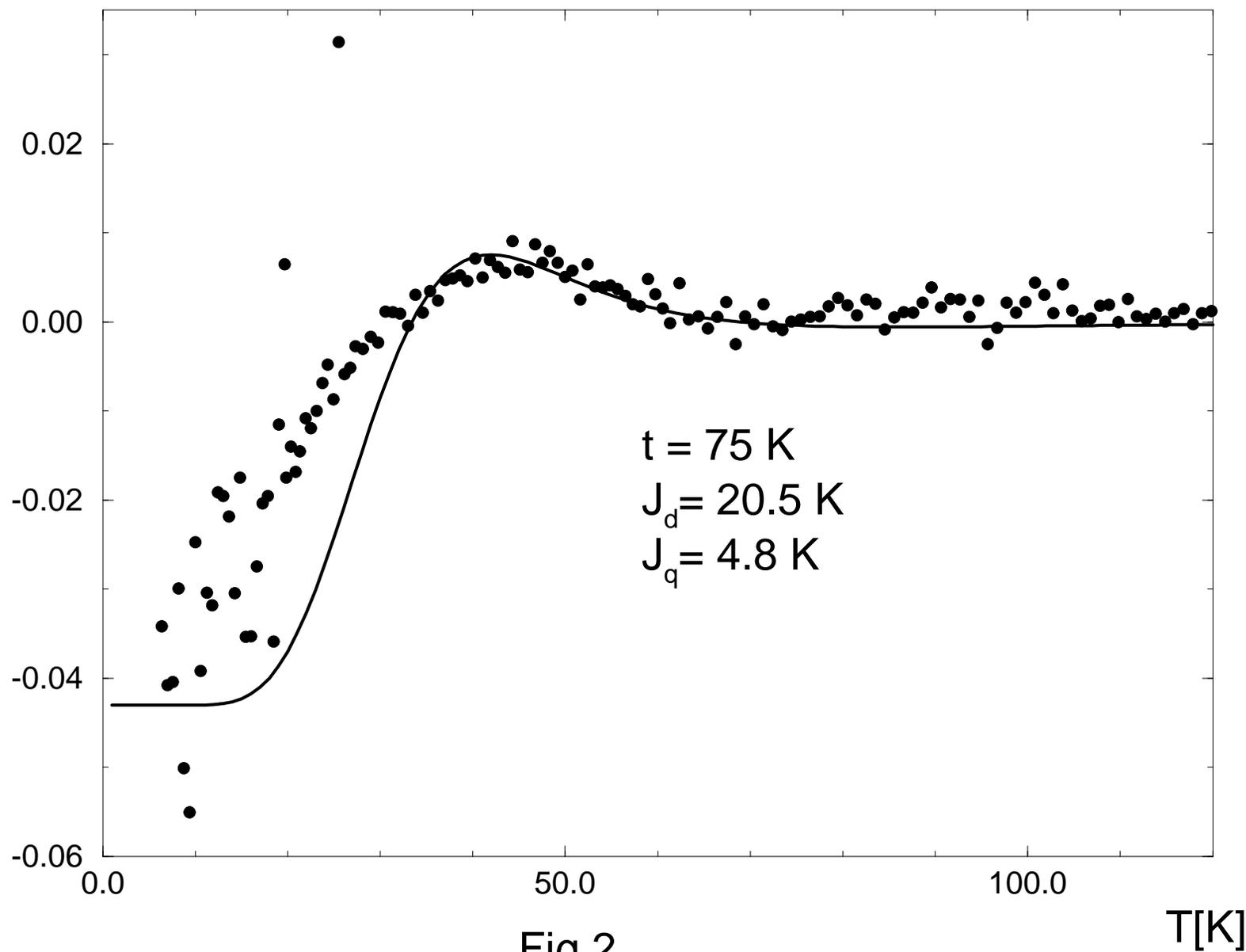
\end{document}